# Electric field induced resistive switching in $M^{3+}_xV_{1-x}O_2$ ($M^{3+}$= $Ga^{3+}$, $Al^{3+}$) single crystals at temperatures below the T→M2 phase transition.

Larisa Patlagan[1], George M. Reisner[1], Shani Neyshtadt-Ronel[2], Yoav Kalcheim[2] and Bertina Fisher[1*]

[1]Physics Department, Technion, Haifa, 3200003, Israel

[2]Materials Science and Engineering, Technion, Haifa, 3200003, Israel

**Abstract**

The phase diagram of $VO_2$ strained or doped with several trivalent ions consists of four phases; in order of increasing temperatures, three (M1, T and M2) are insulating while the fourth (R), above ~340 K, is metallic. These phases and the three phase transitions have been thoroughly investigated for about half a century by a wide variety of techniques, including electronic transport. While an upwards jump of the resistance of up to a factor of 2 was observed at the T→M2 transition and a drop of several orders of magnitude was observed at the M2→R one, resistive switching at the M1→T transition remained elusive over all these years. Here we report on the investigation of Ga- and Al-doped $VO_2$ single crystals, following the rather surprising appearance of a small and steep drop of a factor of ~ 0.12 in the resistance of Ga-doped $VO_2$ single crystals detected by pulsed and DC I-V measurements carried out at room temperature, below the T→M2 phase transition. Similar results were obtained also from measurements on Al-doped $VO_2$ single crystals. Raman spectra of Ga-, and Al-doped crystals measured as function of temperature are consistent with earlier results obtained from strained or $M^{3+}$-doped $VO_2$. The accumulated results of the IV measurements on Ga-, and Al-doped single crystals provide evidence for identifying the resistive switching at $T_{RS}<T_{T→M2}$ with the M1→T transition.

*Corresponding author: phr06bf@physics.technion.ac.il

## 1. Introduction

The phase diagram of pure $VO_2$ consisting of a low temperature insulating phase of monoclinic structure (M1) and a high temperature metallic phase of tetragonal (rutile- R) structure, is enriched by two additional insulating phases when $VO_2$ is strained[1-2], doped with $Cr$[3-4], $Al$[5-7], $Mg$[8], $Ga$[9-10] or, oxygen defficient.[11] The three insulating phases below the insulating-metal transition (IMT) - $T_{IMT}$~340 K (depending on dopant and its concentration) are best illustrated by chains of the vanadium (V) ions: the monoclinic (M1) phase with chains of tilted V ions, the triclinic (T) phase with two types of chains in which the V ions are paired and tilted to different degrees and the monoclinic (M2) phase with two types of V chains, one in which the V ions pair but do not tilt and another in which the V ions tilt but do not pair. While the T→M2 and M2→R transitions were detected by a wide variety of techniques (thermodynamic, X-Ray, magnetic susceptibility, NMR, Raman spectroscopy and electronic transport), the M1→T transition in strained or $M^{3+}$-doped $VO_2$ was detected only by the occurrence of structural distortions.[3,10] This was justified by the assumption that the T phase is merely a slight structural distortion of the M1 phase, consistent with a continuous crossover from the M1 to the T phase.[3] The spectra of XRD of strained or $M^{3+}$-doped $VO_2$ measured below the T→M2 transition consist of lines of the M1 phase superimposed on lines of the T phase. Such examples are seen in Reference (12) reporting on the Insulator-Insulator (the T→M2) and the Insulator-mixed Metal-Insulator-transitions in single crystals of $Al_xV_{1-x}O_2$ (0.007≤x≤0.02) and their visualization (see powder patterns in Figures 1 and 1S for x=0.02 and x=0.07, respectively).

Pulsed and DC I-V measurements aimed at separating non-thermal from thermal electric field effects were recently carried out on pure and doped $VO_2$ single crystals.[13] In the course of measuring pulsed I-V at, or below room temperature (RT), on a number of $VO_2$ single crystals doped with 0.4% Ga (nominal composition), drops of the resistance ($\Delta R/R \sim 0.1$) were detected. At fixed currents, these, rather small drops rode on the natural drop in resistance with increasing pulse duration due to self-heating (Joule heating). Various possible experimental errors were invoked before such small drops were detected again in additional crystals and regarded as a genuine phenomenon. Reported herein are results from measurements on Ga-, and Al-doped $VO_2$ single crystals that provide evidence that these drops represent resistive switching (RS) related to the M1→T transition. Raman spectra of a Ga-, and an Al-doped crystal resolved their structures as function of temperature above RT.

## 2. Experimental

Millimeter size Al-, and Ga-doped $VO_2$ single crystals were grown by the self-flux method from high purity $V_2O_5$ powder mixed with $Al_2O_3$, or $Ga_2O_3$ powder corresponding to the nominal composition of $M_xV_{1-x}O_2$ (M=Ga, Al), in flowing Nitrogen. The crystals are fragile and tend to crack upon temperature or I-V cycling through IMT. I(V) and R(T) of the crystals were measured in the two-probe configurations (with connected adjacent voltage and current probes). The contacts were indium-amalgam dots. These are ohmic contacts of low resistance for insulating (semiconducting) $VO_2$. With these contacts, the samples are free to move, being held only by surface tension. R(T) and DC I-V were measured using a YEW type 3036 X-Y recorder. The pulsed I(V) measurements were carried out using a Keithley 237 source and a Keysight DSOX 2002 oscilloscope. A test resistor R1, stable at high currents, was connected in series with the crystal. Raman analysis was performed using Witec alpha300 microscope, using WiTec green laser (532 nm) set to 1mw. The sample was placed on a sapphire substrate using a Kapton tape. The substrate itself was taped to the bottom of the Linkam DSC600 temperature-controlled stage using an additional Kapton tape to allow good thermal conductivity. The samples were heated from room temperature until the Raman signal of $VO_2$ disappeared (indicating the transition to the metallic phase). The spectra were recorded using a 1200gr/mm grating centered at 620 $cm^{-1}$.

## 3. Results and discussions.

### A. Ga-doped single crystals.

As mentioned in the Introduction, we started recording the steep drops in the V(t) pulses of the Ga-doped $VO_2$ single crystals only after several such drops reappeared in additional crystals thus convincing us that they represent a genuine effect. Figure 1(a) shows the first V(t) pulses of 2 ms durations recorded from the screen of the oscilloscope for two applied currents on crystal $Ga_{0.004}V_{0.996}O_2$ (8) at ambient temperature $T_0$=275 K (green trace). V1 (yellow trace) is the voltage drop on the small test resistor in series with the crystal – R1 and V2 is the total voltage. Note that the ratio of the scales of V2 and V1 is 40. For I=10.1 mA the main jump occurred at t=1 ms while for the higher current, I=10.5 mA, it occurred earlier, at t=0.25 ms. For a jump representing RS at a fixed $T_{RS}$, this temperature is reached at shorter times for higher currents (higher Joule power).

Figure 1(b) shows all data points forming the I(V) traces for pulses of durations τ=1 and 2 ms. Traces were added also for intermediate times of 0.25, 0.5 and 0.75 ms. All nonlinear I(V) traces coincide for V≤40 V; those for t ≤ 1ms coincide for V ≤ 70 V; above this voltage separate, nonlinear traces showing similar small jumps appear at lower currents at the ends of higher pulse durations. Since nonthermal electric field effects are faster than the thermal ones, the traces corresponding to different pulse durations separate when the thermal effects become dominant. The traces shown in Fig. 1(b) were recorded in the thermal range of I(V) and τ=2 ms; the onsets of RS are marked by red symbols. The DC I(V) at the left of this Figure, representing steady state, is strongly nonlinear. The voltage reaches a maximum $V_{max}$=24.5 V around I=4.3 mA; the decrease from the maximum marks the onset of current controlled negative differential resistance (CCNDR).[14,15] A clear mark of the T→M2 transition (a sudden increase of V) is seen at I=6.2 mA.[9] Finally, the transition to the mixed insulator-metal phase (M2→R) is seen around I=9 mA. To protect the crystal from cracking, tracing was stopped at the onset of the mixed state, but this precaution was not successful. The fast drop of the current on the way back (dashed trace) was due to a crack. Fortunately, the crack was close to one of the contacts and could be repaired, allowing the measurement of R(T).

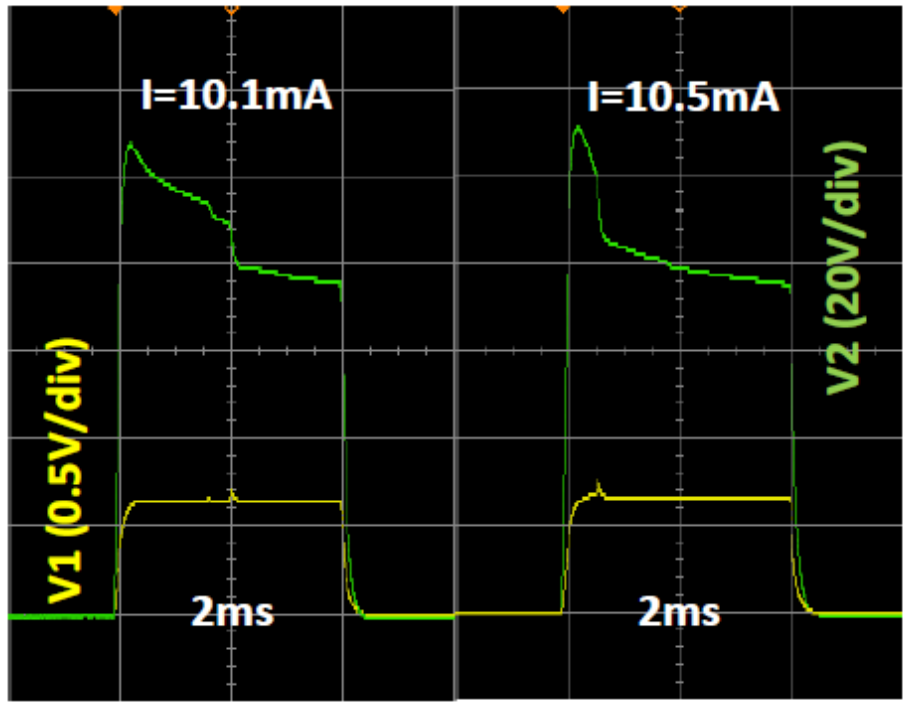

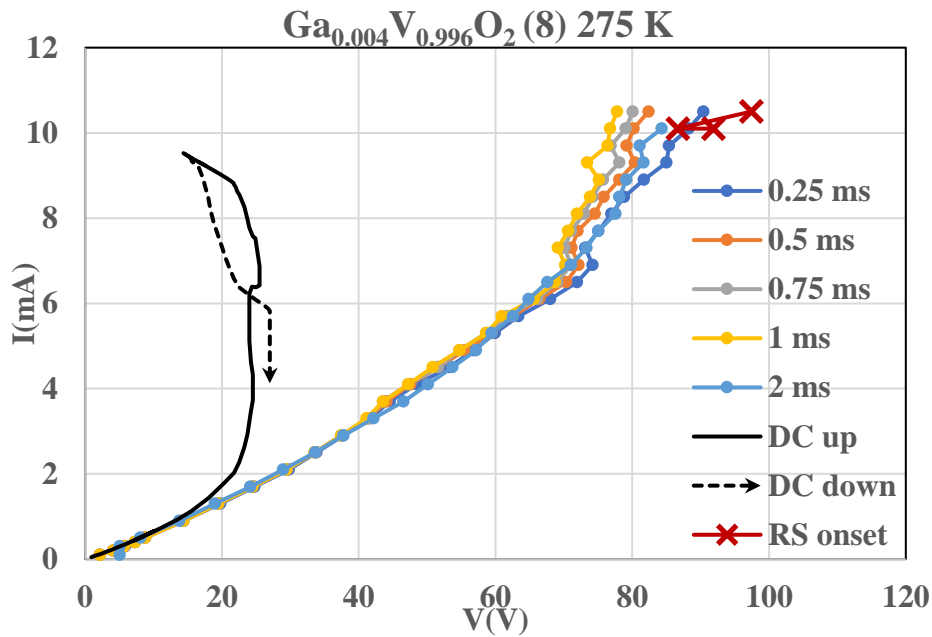

Fig.1. (a) V(t) for two current pulses of 2 ms durations applied on crystal $Ga_{0.004}V_{0.996}O_2$ (8); (b) Pulsed and DC I-V characteristics for this crystal at ambient temperature of 275 K. Data points represent the end of the pulses and intermediate times: t=0.25, 0.5 and 0.75 ms, and the onsets of the jumps from (a). The mark of the T→M2 transition and that of the onset of the mixed insulator metal phase are seen on the DC I(V) trace for increasing current (DC up). The sharp drop of the current on the DC down trace is caused by cracks.

Pulsed measurements were repeated on crystal $Ga_{0.004}V_{0.996}O_2$ (9) from the same batch. Figure 2(a) shows V(t) pulses of 64 ms durations recorded from the screen of the oscilloscope for four applied currents on crystal $Ga_{0.004}V_{0.996}O_2$ (9) at ambient temperature $T_0$=275 K. A tiny downwards drop appears for 2.4 mA at t=24 ms followed by a second one for 2.5 mA at t=19 ms. Several V(t) traces with evidence for the T→M2 transition were recorded, two of them for I=3 mA and 3.3 mA are shown in Figure 2(a). Here again the discontinuity in V(t) appears earlier for higher currents. While the tiny downwards drops attributed to the new RS are steep, the increasing V(t) riding on the heating trace, consistent with the T→M2 transition, is smeared; this does not contradict T→M2 being a first order phase transition. In the I-V characteristic driven by self-heating the T→M2 transition induces a voltage controlled NDR (VCNDR) which requires constant voltage conditions for correct display;[14] the smearing is caused by our pulse generator that provides constant current conditions.

Figure 2(b) shows all data points representing the I(V) traces for pulses of various durations from τ=1 to 64 ms. The onsets of RS representing the drops in V(t) in Fig,2(a) marked by red symbols-at, and those representing the T→M2 transition, at higher currents (higher temperatures), by purple symbols. The sequence of the power (=VI) for two types of RS switching, are consistent with the order of the M1→T and T→M2 transition temperatures.

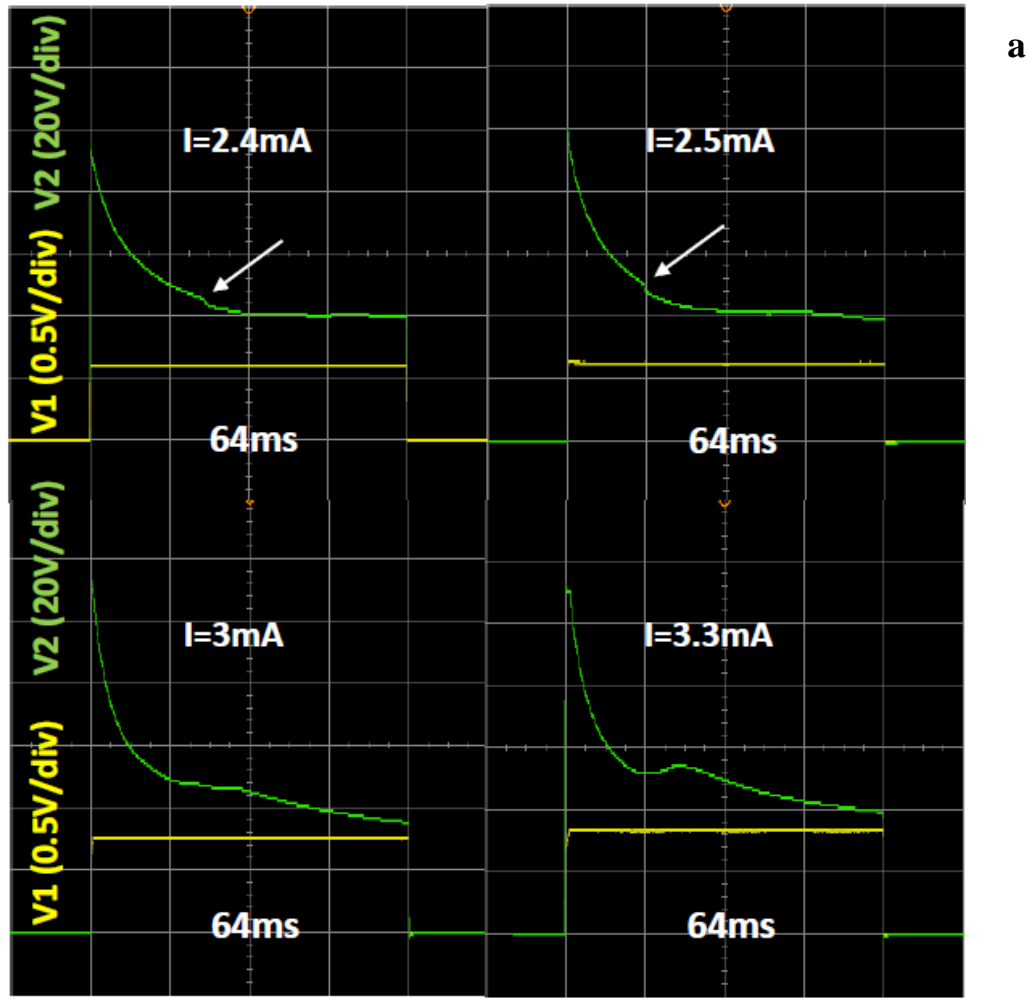

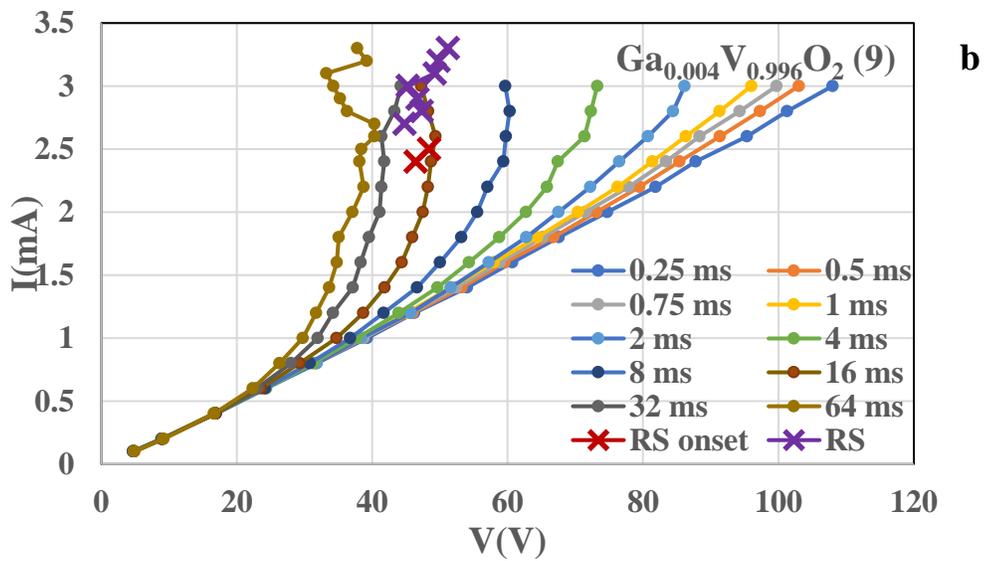

Fig.2 (a) V(t) for four current pulses of 64 ms durations applied on crystal $Ga_{0.004}V_{0.996}O_2$ (9) at 275 K; (b) pulsed I-V characteristics for this crystal at ambient temperature of 275 K. Data points represent the end of the pulses and intermediate times: t=0.25, 0.5 and 0.75 ms and the onsets of the jumps (red and purple marks) from (a).

R(T) and DC I-V measurements were carried out on a third crystal from the same batch, $Ga_{0.004}V_{0.996}O_2$ (11). Figure 3(a) shows the semi log plots of R versus 1000/T for this crystal between 275 K and 359 K, upon heating and cooling. R(T) of this crystal exhibits two activated ranges of practically equal activation energies of 0.27 eV with a small jump corresponding to the known T→M2 transition around 307 K and the prominent IMT at 355 K and MIT at 337 K. No mark of a transition is detected on the R(T) above 275 K and the T→M2 transition.

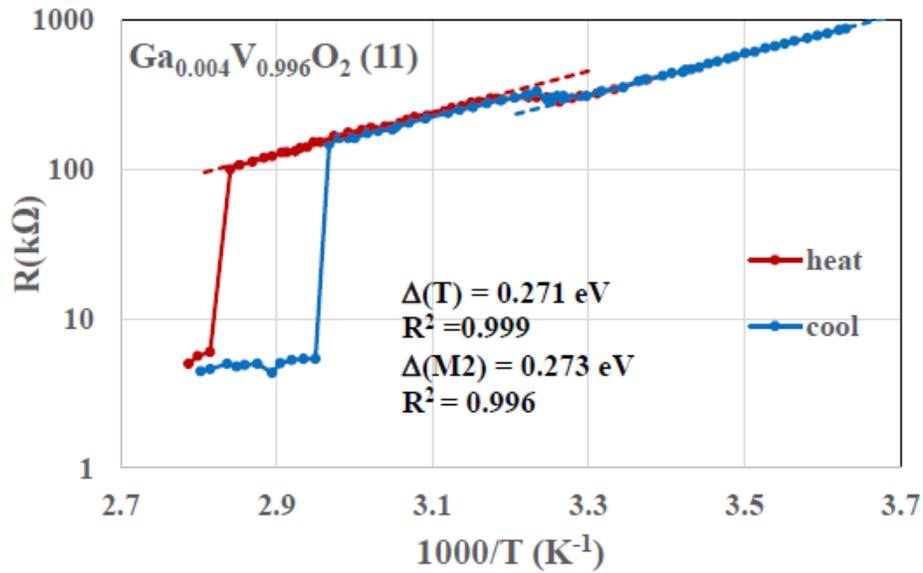

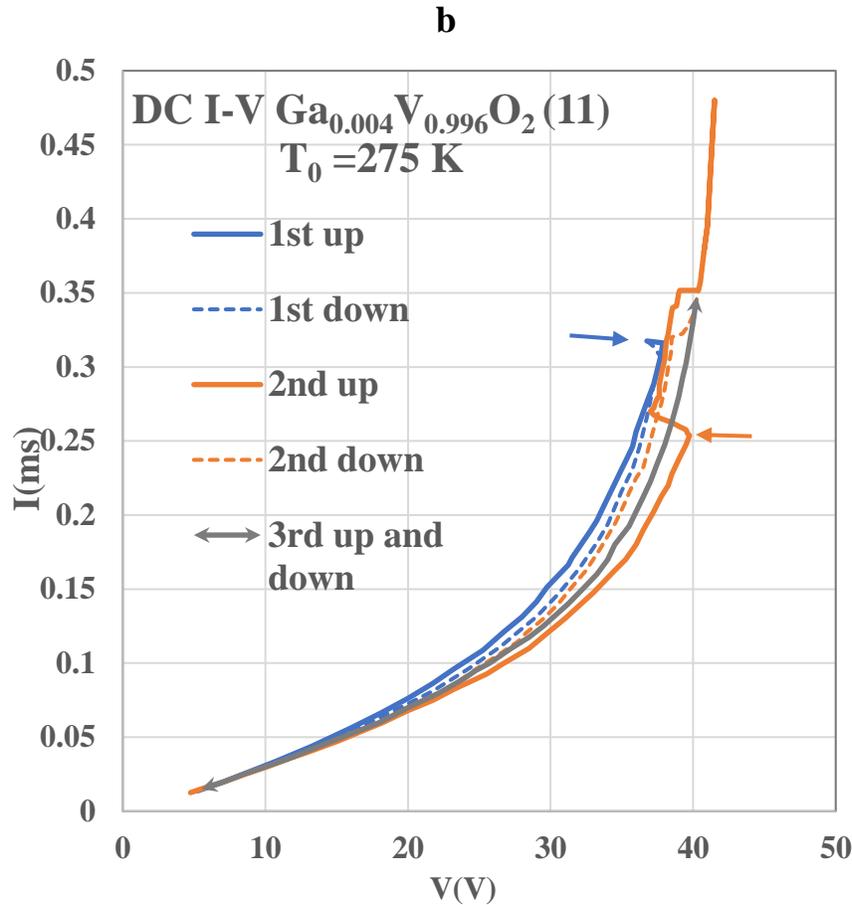

Figure 3. (a) Semi-log plot of R versus 1000/T measured on crystal $Ga_{0.004}V_{0.996}O_2$ (11). (b) Three consecutive I-V loops measured on crystal $Ga_{0.004}V_{0.996}O_2$ (11) at ambient $T_0$=275 K.

Pulsed I-V measurements on crystal $Ga_{0.004}V_{0.996}O_2$ (11) using our pulse generator were excluded by its high resistance even after decreasing its length by a factor of 2. Figure 3(b) shows three consecutive loops of DC I-V measured on this crystal. DC IV measurements represent steady state[12] which is a great advantage over pulsed measurements (see below).

Upon increasing current a small decrease in V (decrease in R), resembling the RS in the pulsed measurements before, appeared at the upper end of the "1st up" trace (I=0.32 mA). The "1st down" trace starts from the end of the jump, crosses the "1st up" trace below I=0.3 mA, continues its decrease at higher voltages than the up-trace and coalesces with it towards its lower end. This behavior is in contrast to thermal hysteresis (always present in DC IV) which would have kept the return path on the smaller voltage (smaller resistance) side.

The load resistance was decreased to allow an increased range of voltages. A prominent RS appeared upon increasing current of the "2nd up" loop around I= 0.25 mA; here the change in R was $\Delta R/R \approx -0.2$. This transition was followed by a modest increase in voltage for I=0.35 mA (corresponding to $\Delta R/R \approx +0.02$) that marks the T→M2 transition and continued to increase nonlinearly. The "2nd down" trace exhibited the thermal hysteresis of the M2→T transition but showed no sign of RS until it coalesced with the previous traces at low currents. Along the "3rd" closed loop no RS nor hysteresis are seen; it turned into a single (gray solid) line that joins the lower common trace of the three loops with the upper portion of the 2nd loop. The two RS seen on the same DC I-V trace (2nd up) appear at close values of Joule power

(P=IV); from the data on this trace one finds that $P_{T\to M2}/P_{M1\to T}=1.36$. In a first approximation, at steady state, P is proportional to $T-T_o$. From Figure 3(a), $T_{T\to M2} \approx 307$ K. and thus, $(307-275)/(T_{M1\to T}-275)$ =1.36. From this simple relation the estimated $T_{RS}\approx 298$ K. Note that the two transitions are separated by ~11 K.

New, interesting evidence of this RS was uncovered by the investigation of crystal $Ga_{0.004}V_{0.996}O_2$ (12) (see Figure 4). Figure 4(a) shows two traces of R(T) measured with low DC current upon cooling from RT to 275 K, on the virgin crystal (1st cool) and before a second stage of measurements (2nd cool). The 1st trace shows a moderate drop in R(T) starting at ~ 288 K and a moderate rise in R(T) starting at ~280 K (see arrows) riding on an increasing R with decreasing T. This seems to hint towards the two events (the M2→T and the T→M1 transitions) separated by 10 K.

Figure 4(b) shows pulse and DC I-V measurements that followed the 1st cooling. I-V measurements were carried out using pulses of 1 ms durations and, as before, data for intermediate times (0.25 - 0.75 ms) were also recorded. The I(V) plots are strongly non-linear and no events were detected during the durations of the pulses, except for low downwards slopes, up to about V=35 V and large slopes at higher voltages. With increasing currents, the highest voltage reached was 50 V at 0.25 ms (and lower for longer durations) above which V exhibited large drops corresponding to large drops in resistance, indicating the onset of the mixed metal insulator (mMI) state. This set of measurements was followed by the DC I-V measurements (see left axis). While crystal $Ga_{0.004}V_{0.996}O_2$ (11) showed RS upon the 1st and the 2nd current increase but no RS on the way back, the DC I-V loop in crystal $Ga_{0.004}V_{0.996}O_2$ (12) exhibited jumps upon increasing as well as upon decreasing currents, including hysteresis. ***This loop contains the mark on a DC I-V of a first-order phase transitions, upon heating and cooling.*** Thus, it is conceivable to identify the jumps on this trace with the elusive resistive switching of the M1→T transition. Figure 4(c) shows R(V) calculated from the data in (b). Upon increasing voltage, it exhibits a drop of $\Delta R/R > 0.2$.

This sample underwent slow heating to RT (overnight). The trace of the 2nd cool (axis at right) in Figure 4(a) shows that the new resistance of the crystal was lower by more than a factor of 2 than that of the first cooling with poorly identifiable marks of transitions. Repeated DC I-V characteristics appeared at currents lower than the first one (higher resistances) and showed only faint marks that can be associated with the two adjacent phase transitions, but the return path was smooth as in Figure 3(b). Further attempts to reproduce a DC loop similar to that in Figure 4 (b), in this crystal, or in other Ga-doped crystals, failed.

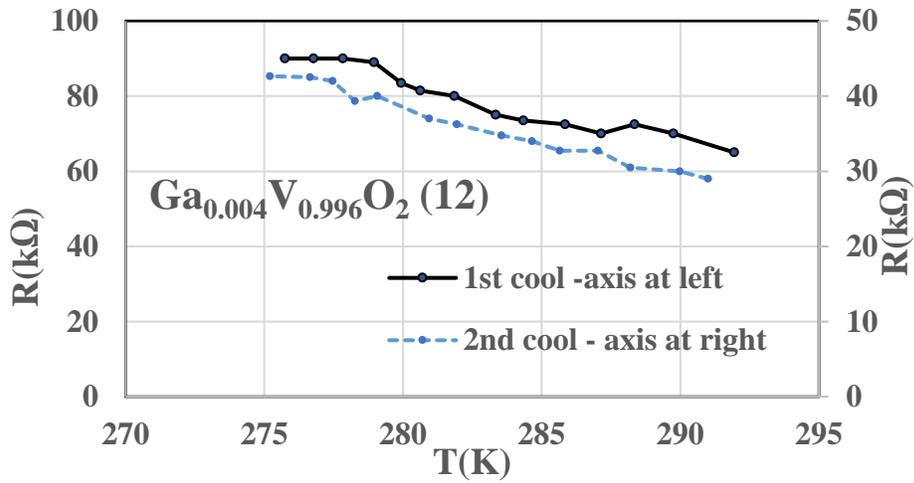

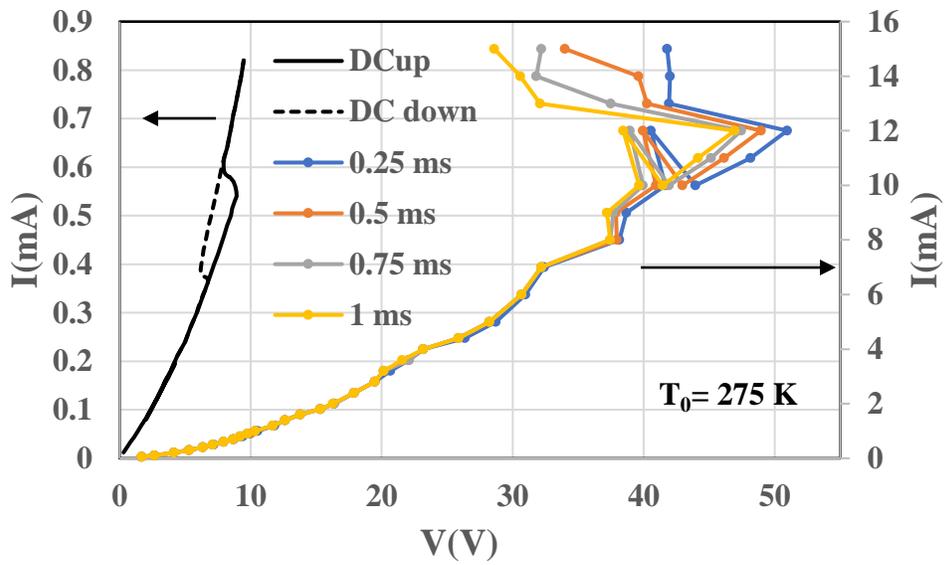

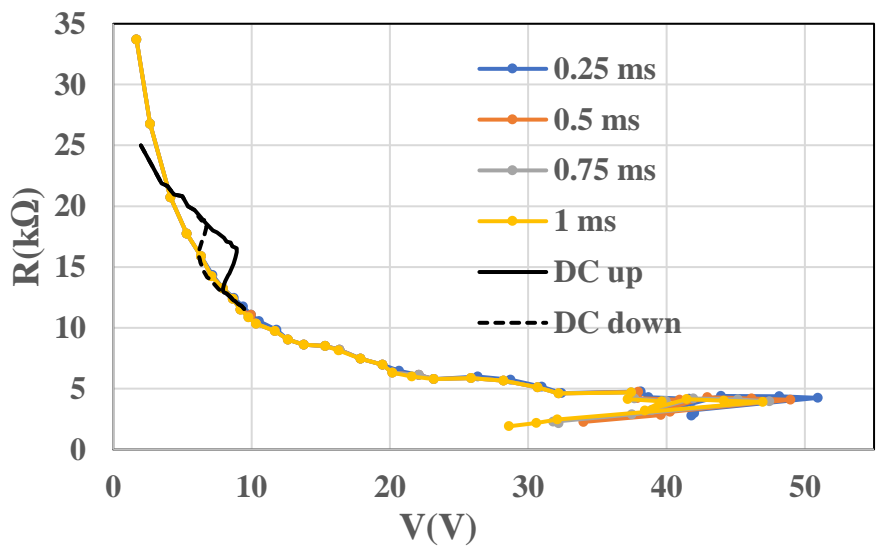

**Figure 4.** (a) R(T) measured on crystal $Ga_{0.004}V_{0.996}O_2$ (12) upon cooling from RT to 275 K. (b) I-V pulsed I-V characteristics measured after the first cooling (right axis) and DC I-V (left axis). (c) R(=V/I) calculated from (b).

B. Al doped crystals

The above investigation was extended for Al-doped single crystals. At this stage we aimed at obtaining the marks of RS, of T→M2 and of I→mMI transitions along the same DC I-V loop. This would allow us to estimate with confidence $T_{M1→T}$. After a series of failures, we obtained the desired results for crystal $Al_{0.007}V_{0.993}O_2$ (10). In this case the I-V measurements preceded the R(T) ones.

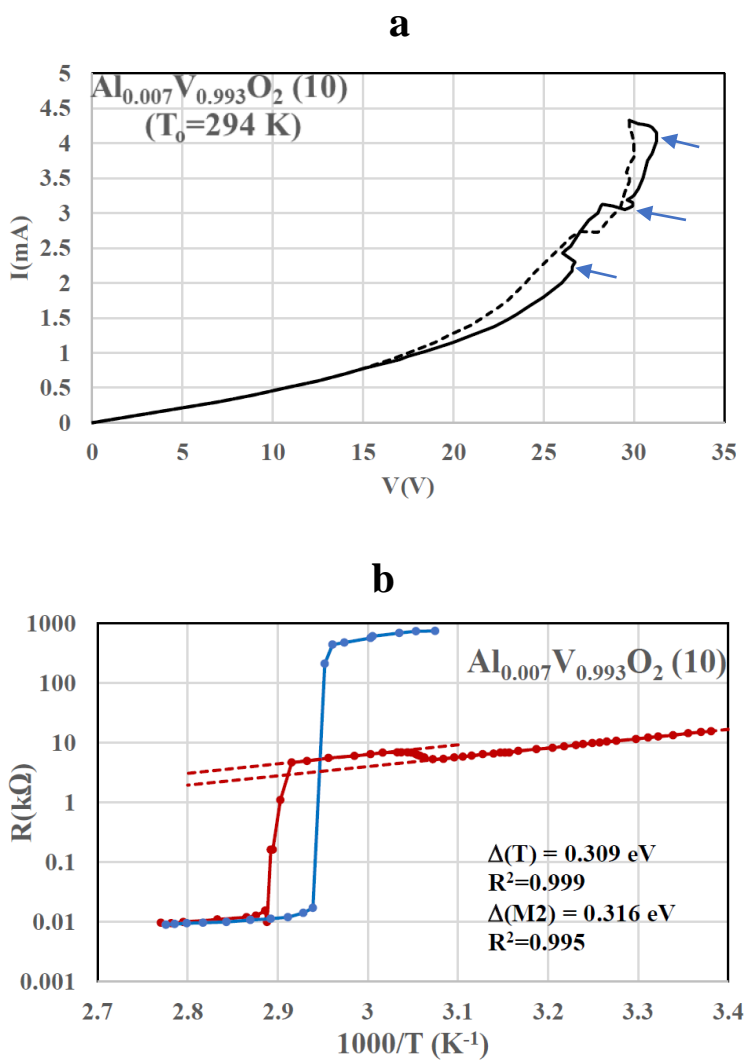

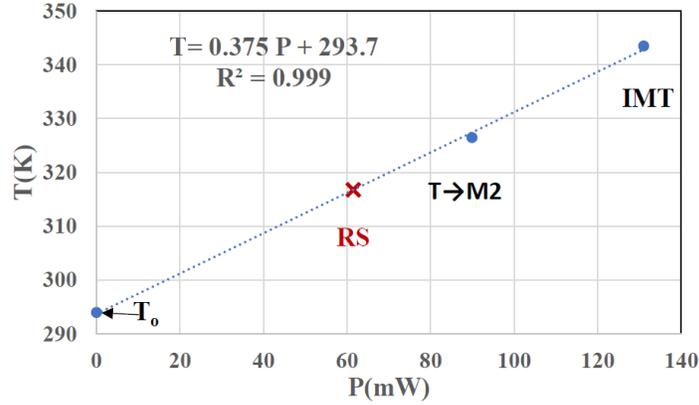

Figure 5. (a) DC I-V loop measured on crystal $Al_{0.007}V_{0.993}O_2$ (10) at ambient $T_0=294$ K. (b) semi-log plot of R versus 1000/T measured on crystal $Al_{0.007}V_{0.993}O_2$ (10). The three blue arrows along the up trace show the marks of the new RS, T-M2 and I→mMI transitions. (c) Estimating $T_{RS}$ from the T and P data for T→M2 and IMT and P for RS (from (a).

Figure 5 represents the results obtained from I-V and R(T) measurements carried out on crystal $Al_{0.007}V_{0.993}O_2$ (10). The I-V characteristics in Figure 5(a) is nonlinear due to Joule heating. The three blue arrows show the marks of the RS, the T→M2 and the I→mMI transitions that occur upon increasing current (solid line). Upon decreasing current (dashed line) only the marks of mMI→I and of M2→I transitions are seen shifted by a large hysteresis. At the end of several, reproducible I-V characteristics the sample broke into two. R vs 1000/T shown in Figure 5(b) represents the results obtained from about half of the initial crystal. As before, R(1/T) shows two ranges of activated conduction with close activation energies typical for Al-doped crystals. Upon cooling, at the MIT the resistance jumped by about two orders of magnitude, the crystal cracked. Two important data are provided by the heating trace: the onsets of the T→M2 transition ($T_{T→M2}=326.5$ K) and of IMT ($T_{IMT}=343.5$ K), These temperatures as function of the corresponding data of P obtained from (a) and T=294 K for P=0, were plotted in Figure 5(c). A linear trendline corresponding to the assumption that in steady state P(T) is proportional to $(T-T_0)$ fits almost perfectly the three data points (see trendline label on this Figure). This adds validity to the assumption used in the previous section. The power P=61.5 mW at the onset of RS in (a) corresponds to $T_{RS}=316.8$ K. $T_{RS}$ and $T_{T→M2}$ are separated by ~ 10 K.

To sum up the experimental results of the electrical measurements:

The resistive switching (RS) associated with the M1→T transition obscured for decades, has surprisingly appeared in pulsed I-V characteristics measured on Ga-doped single crystals. While in pulsed measurements RS can be detected only upon (Joule) heating, DC I-V allows heating upon increasing currents as well as cooling upon decreasing currents under steady state conditions. Extending the search for this RS for Ga-, and Al-doped crystals to DC I-V measurements it was found that while RS appears eventually upon increasing current, in only **one case of many**, jumps appeared upon increasing as well as upon decreasing currents (see Figure 4(b)). It is notable that in that unique case a hint of the T→M1 transition appeared on the trace of R(T) upon cooling from RT (see Figure 4(a)).

Phase transitions occur by nucleation and growth. The bottleneck of the event dealt with in literature (see e.g. Reference[16]) is the formation of a proper nucleus that allows the growth of the second phase within the initial one. With no nucleation seeds introduced intentionally, the present results indicate that the nucleation seeds of the M1→T transition in the Ga-, and Al-doped single crystals are hot spots selected by self-heating which is sensitive to small deviations from uniformity of the resistivity on the microscopic scale. The scarcity of events related to the opposite transition, T→M1, leaves no indications as to the nucleation seeds that enhance this transition. It is suggested that these may be small domains of the M1 phase left within the T phase from the previous transition.

### C. Raman spectra of Ga- and Al-doped VO$_2$ single crystals.

Raman spectra were measured on a Ga$_{0.004}$V$_{0.996}$O$_2$ crystal, an Al$_{0.007}$V$_{0.993}$O$_2$ crystal and a pristine VO$_2$ single crystal used for comparison. The Raman spectrum for the Ga$_{0.004}$V$_{0.996}$O$_2$ (9), agrees well with the findings of the electrical measurements.

The Raman spectrum of pristine VO$_2$ (blue line in figure 6(a)) exhibits 3 main peaks, at 192 cm-1, 223 cm-1 and approx. 610 cm-1, that are typical to VO$_2$ in the crystalline structure. These peaks are attributed to the V-V vibration along the C$_R$ (C-axis of the rutile structure), V-V vibration perpendicular to C$_R$, and the V-O vibration, respectively.[8] The positions of these peaks in the pristine sample are marked with blue dashed line. Additional peaks at 306, 390 and 500 cm-1 are also typical to the VO$_2$-M1.[17]

The V-O and V-V frequencies indicate the predominant phases of the VO$_2$. The frequencies of VO$_2$-M2 phase are shifted to 640-650 & 200 cm-1, respectively. For the T-phase, described often as a combination between M1 and M2, the V-O peak is split into 2 peaks, 615 cm-1 and 585 cm-1, while the V-V peaks remain similar to M1.[17]

The RT spectrum of Ga doped VO$_2$ is in grey line in Figure 6(a) and blue line in figure 6(b). Compared to the known spectrum of VO$_2$(M1) the V-0 and the V-V peaks are shifted to 632 and 200 cm$^{-1}$, respectively, being closer to the expected M2-VO$_2$ spectrum. Some splitting in the VO$_2$ peak, that could be seen in the spectrum may imply on the existence of mixed phases (both M2 and T), however definite identification is difficult.

Phase behavior of doped VO$_2$ as a function of temperature could be tracked using Raman measurements performed in a temperature-controlled stage, as could be seen in figure 6(b). Upon moderate heating to 313 K, the peaks are further shifted to 650 and 202 cm-1 – i.e., the M2 phase is formed as could be seen in Figure 6(b). The T→M2 transition seen in Figure 3(a) for Ga$_{0.004}$V$_{0.996}$O$_2$ +[11] (from the same batch) occurs around 307 K. The Raman spectrum remains unchanged until 358 K, where these peaks disappear indicating the formation of metallic rutile phase. According to Figure 3(a) T$_{IMT}$=355 K.

For the Al- doped sample the main change appears in the V-O peak that is split into 2 peaks, at 615 cm-1 and 585 cm-1, as can be seen in the orange line in Figure 6(a) and in the blue line in Figure 6(c)

This split is typical for VO$_2$ in triclinic structure (T)[18], that have been shown for Al doped VO$_2$ by Strelcov er al.[19] Additional peaks around 1000 cm-1 and 285 cm-1 are characteristic to V$_2$O$_5$, powder on the surface of the crystal that was not converted into VO$_2$.[20]

At temperatures above 325 K the V-O peak shifts to higher energy, together with the V-V peak, as can be seen in comparison between the orange and the grey line in Figure 6(c). The new peaks, located at 654 cm-1 and 202 cm-1, are typical to the M2 phase of VO$_2$, blue shifted compared to M1.[8] Upon further heating, between 343 and 348 K all the VO$_2$ peaks disappear and only the V$_2$O$_5$ peaks remain, as could be seen in the light blue line in figure 6(c), indicating transition to the rutile phase at high temperatures. In Figure 3(b) above and S2(a) in the Supplementary material for Al$_{0.007}$V$_{0.993}$O$_2$ (10) and (9) crystals, respectively, T$_{IMT}$=344 and 347 K, respectively.

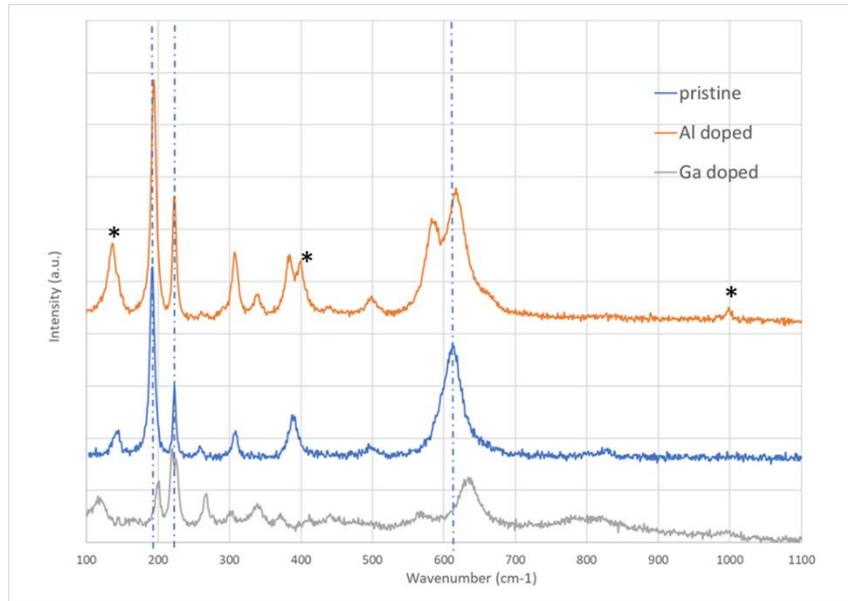

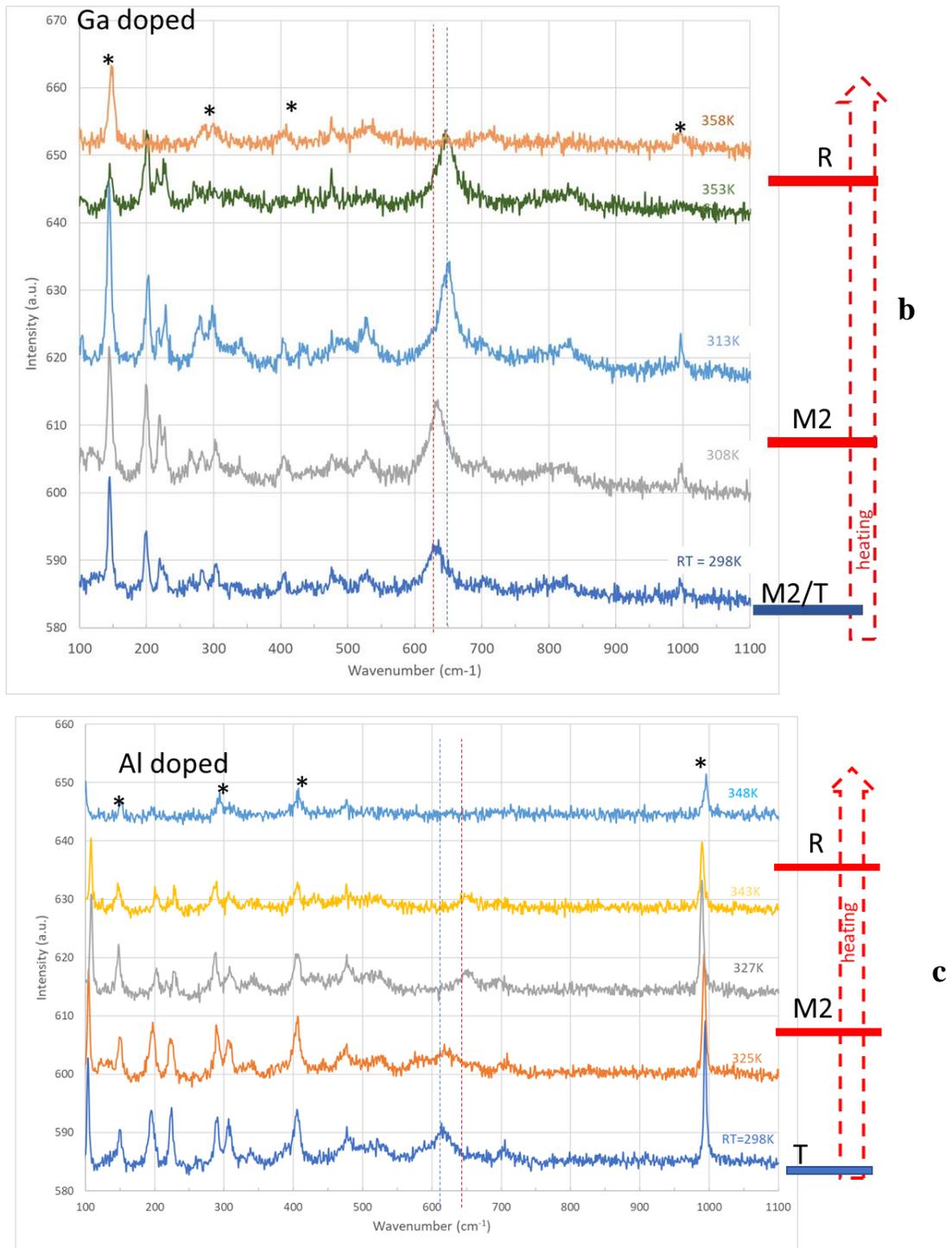

Figure 6(a) - Raman spectrum of Al (orange line) and Ga doped (grey line) VO$_2$ needles, compared to undoped VO$_2$ needles obtained in a reference experiment; (b) Raman spectrum of Ga- doped VO$_2$ during heating experiment; (c) Raman spectrum of Al doped VO$_2$ obtained during heating experiment. In (b) and (c) the positions of the shifted VO$_2$ peaks are marked with blue dashed line for room temperature and red dashed line for high temperature.

In summary, the T→M2 transition and the Insulator Metal-Insulator transition in Ga-, and Al-doped VO$_2$ single crystals found by electrical measurements are in close agreement with the results of Raman spectroscopy. No evidence was found for the M1→T transition in the present Raman spectra measured above RT. Atkin et al.,[17] on Raman carried out above RT on strained VO$_2$, showed that the T- phase emerges as an intermediate structure from a continuous distortion of the M1 lattice. The M2 phase, by contrast, shows a discontinuous transition when emerging from either the M1 or the T phase.

4. Conclusion.

The decades old absence of resistive switching related to the M1→T transition in $M^{3+}_xV_{1-x}O_2$ has now been found in pulsed and DC I-V measurements carried out on Ga-, and Al-doped VO$_2$ single crystals at ambient temperatures slightly below, or at room temperatures. While presenting no candidacy for interesting technical applications, this transition provides an interesting test bed for the study of one of the most intricate problems in materials science: the nucleation and growth of one solid phase within another. The absolute magnitude of the resistive switching related to the M1→T transition is comparable to that of the M2→T transition, but while the latter is perfectly reproducible (including hysteresis) the first can be hardly discerned.

**Supplementary Material**

See supplementary material for:
Additional pulsed I-V measurements on a Ga-doped VO$_2$ crystal
Additional DC I-V measurements on two Al-doped crystals
R(T) of Ga-doped crystals carried out after prior IV measurements.

**Acknowledgements:** We are indebted to I. Taitler for his valuable advice concerning the pulsed measurements. We are grateful to Lior Kornblum and Eilam Yalon for illuminating discussions. We thank Chubin Huang for growing the pristine VO$_2$ crystals. We likewise thank Adi Levi and Ayala Hodara for their assistance in pertinent information search.

The authors declare no conflict of interests.

**References.**